\documentclass[iop,apjl]{emulateapj}

\usepackage{graphicx} 
\usepackage{dcolumn} 
\usepackage{bm} 
\usepackage{amsfonts,amsmath,amssymb,mathrsfs}
\usepackage{color}
\usepackage{epsfig}
\usepackage{natbib,times}
\usepackage{url}
\usepackage{times}

\pdfoutput=1

\shorttitle{Magnetar Formation from BNS mergers}
\shortauthors{Giacomazzo \& Perna}

\begin{document}

\title{Formation of Stable Magnetars from Binary Neutron Star Mergers}

\author{Bruno {Giacomazzo}\altaffilmark{1} and Rosalba
  {Perna}\altaffilmark{2}}

\altaffiltext{1}{JILA, University of Colorado and National Institute
  of Standards and Technology, Boulder, CO 80309, USA}

\altaffiltext{2}{JILA and Department of Astrophysical and Planetary
  Sciences, University of Colorado, Boulder, CO 80309, USA}

\begin{abstract}
  By performing fully general relativistic magnetohydrodynamic
  simulations of binary neutron star mergers, we investigate the
  possibility that the end result of the merger is a stable
  magnetar. In particular, we show that, for a binary composed of two
  equal-mass neutron stars (NSs) of gravitational mass $M \sim 1.2
  M_\odot$ and equation of state similar to Shen et al. at high
  densities, the merger product is a stable NS. Such NS is found to be
  differentially rotating and ultraspinning with spin parameter
  $J/M^2\sim 0.86$, where $J$ is its total angular momentum, and it is
  surrounded by a disk of $\approx 0.1 M_\odot$. While in our global
  simulations the magnetic field is amplified by about two orders of
  magnitude, local simulations have shown that hydrodynamic
  instabilities and the onset of the magnetorotational instability
  could further increase the magnetic field strength up to magnetar
  levels. This leads to the interesting possibility that, for some NS
  mergers, a stable and magnetized NS surrounded by an accretion disk
  could be formed. We discuss the impact of these new results for the
  emission of electromagnetic counterparts of gravitational wave
  signals and for the central engine of short gamma-ray bursts.
\end{abstract}


\section{Introduction}
\label{introduction}
Binary neutron stars (BNSs) are the leading candidates for the central
engine of short gamma-ray bursts (SGRBs; \citealt{1984SvAL...10..177B,
  1986ApJ...308L..43P, 1989Natur.340..126E}). They are also one of the
most powerful sources of gravitational waves (GWs), and advanced
interferometric detectors are expected to observe these sources at
rates of $\sim0.4-400$ events per year~\citep{Abadie:2010}.

Fully general-relativistic simulations have shown how such mergers can
lead to the formation of hypermassive neutron stars (HMNSs), i.e., of
NSs with masses larger than the maximum mass that can be supported by
uniform rotation. Due to loss of angular momentum via GW emission and
magnetic fields, the HMNS eventually collapses in less than $\sim
  1\, {\rm s}$ to a black hole (BH) surrounded by an accretion
disk~\citep{Baiotti08,Kiuchi2009,Giacomazzo2011,Faber2012}. When
magnetic fields are present, they can provide a mechanism to extract
energy from the BH and disk, and power collimated relativistic
jets~\citep{2011ApJ...732L...6R}.

A source of uncertainty in BNS simulations is due to the lack of
detailed knowledge of the equation of state (EOS) of NSs
(see~\citealt{Hebeler2013} for a recent discussion of EOSs). Current
observations have shown that NSs with masses of $\sim 2 M_\odot$
exist~\citep{Demorest2010,Antoniadis2013}, and therefore the NS EOS
must support a mass at least as large as that. This opens the
interesting possibility that the merger of BNSs could produce not only
HMNSs, but also NSs which are stable against gravitational
collapse. This possibility has interesting applications for
observations of SGRBs~\citep{dai2006,Belczynski2008,Rowlinson2013}, as
well as for the possible emission of long periodic GW
signals~\citep{stella2009} and their electromagnetic
counterparts~\citep{Zhang2013,Gao2013,Fan2013}. In particular,
\citet{Rowlinson2013} found that several of the SGRBs that show a
plateau phase in the X-ray lightcurve may be explained with the
formation of a stable magnetar after the BNS merger. X-ray lightcurves
indeed suggest a long-lived central engine~\citep{Margutti2011}, which
is hard to explain if all the BNS mergers produce an HMNS which
collapses to a BH in less than $\sim 1 {\rm s}$; the remnant torus is
also expected to be completely accreted on the same
timescale~\citep{2011ApJ...732L...6R}, unless magnetic or
gravitational instabilities set in~\citep{proga06,perna06}. The
formation of a stable magnetar could also explain the extended
emission observed in several SGRBs~\citep{Metzger2008}.

In this Letter we investigate, for the first time and in fully
general-relativistic MHD, the regime in which BNSs may lead to the
formation of a stable NS. By considering models with and without
magnetic fields, and by performing the longest (to date) general
relativistic simulations of magnetized BNS mergers, we show that
stable magnetized NSs can indeed be formed for some range of masses,
and we discuss the implications of these new results for the
electromagnetic signals that they may emit. Note that the
  formation of a stable magnetar is a non-trivial outcome; in fact, even
  if the total mass of a BNS system is below the maximum mass,
  collapse to BH may still happen if the central density of the merger
  product is above a certain value (depending on the EOS). NSs with
  masses below the maximum mass can indeed collapse to BHs (e.g.,
  model D0 in~\citealt{Baiotti2007}), and the stability properties of
  NSs are defined by both their masses and central
  densities~\citep{Friedman1988}. In this Letter, together with presenting the
  first simulation of the formation of a stable NS from a binary merger, we
  also discuss the NS final spin, the formation of a disk, magnetic field
  amplification, the GW signal, and possible electromagnetic counterparts.

\begin{table*}[t!h!]
  \begin{center}
  \caption{Initial Data\label{table1}}
  \begin{tabular}{lcccccccc}
    \tableline\tableline 
    {Binary} &
    {$\rho_{\rm max}~({\rm g\, cm^{-3}})$} &
    {$M^{\infty}_{1,2}~(M_{\odot})$} &
    {${\cal C}^{\infty}_{1,2}$} &
    {$d/M_{\rm ADM}$} &
    {$M_{\rm ADM}~(M_{\odot})$} &
    {$J~({\rm g}\, {\rm cm}^2\, {\rm s}^{-1})$} &
    {$\Omega_0~({\rm rad\, ms^{-1}})$} &
    {$B_{0}~({\rm G})$}\\
    \tableline
    B0        & $4.40 \times 10^{14}$  & $1.22$ & $0.13$ & $15.6$ & $2.42$ & $5.42 \times 10^{49}$ & $1.70$ & $0$\\
    B12       & $4.40 \times 10^{14}$  & $1.22$ & $0.13$ & $15.6$ & $2.42$ & $5.42 \times 10^{49}$ & $1.70$ & $1.0 \times 10^{12}$\\
    \tableline                                                                                          
  \end{tabular}
  \tablecomments{From left to right, we indicate:  name of model,
     initial value of the maximum density $\rho_{\rm max}$, 
    gravitational mass $M^{\infty}_{1,2}$ of the two NSs 
    and their compactness ${\cal C}^{\infty}_{1,2}$ when at infinity,
    proper initial separation normalized to the initial
    gravitational mass of the binary $M_{\rm ADM}$, initial value
    of the total angular momentum $J$, initial orbital frequency
    $\Omega_0$, and initial maximum value of the magnetic field
    $B_0$.}
  \end{center}
\end{table*}

Section~\ref{methods} details our numerical methods and the initial
models.  Section~\ref{dynamics} describes evolution and dynamics of
these systems, while in Section~\ref{GWs} we discuss their
gravitational and electromagnetic signals. Section~\ref{conclusions}
summarizes our main results. For convenience, we use a system
of units in which $c=G=M_\odot=1$, unless explicitly stated otherwise.

\section{Numerical Methods and Initial Data}
\label{methods}
The simulations presented here were performed using the publicly
available Einstein Toolkit~\citep{Loeffler2012}, coupled with our
fully general relativistic MHD code
\texttt{Whisky}~\citep{Giacomazzo2007,Giacomazzo2011}. Details
  about the numerical methods can be found
  in~\citet{Giacomazzo2011}, except that in our work the spacetime evolution
  is obtained using the \texttt{McLachlan} code~\citep{Loeffler2012},
  and \texttt{Whisky} now implements the modified Lorenz gauge to
  evolve the vector potential and the magnetic
  field~\citep{Farris2012}.
The simulations presented here use adaptive mesh refinement
with six refinement levels; the finest grid covers completely each of
the NSs during the inspiral and merger, while the coarsest grid
extends up to $\sim 777 \,{\rm km}$. Our fiducial runs have a
resolution of $\sim 225 \,{\rm m}$ on the finest grid, but convergence
tests have been performed using both a coarser ($\sim 360 \,{\rm m}$)
and a finer resolution ($\sim 180\, {\rm m}$).

The initial data were produced using the publicly
available code {\tt
  LORENE}~\citep{Taniguchi02b}.\footnote{http://www.lorene.obspm.fr} The
initial solutions for the binaries were obtained assuming a
quasi-circular orbit, an irrotational fluid-velocity field, and a
conformally-flat spatial metric. The matter is modeled using a
polytropic EOS $p = K \rho^{\Gamma}$, where $p$ is the pressure,
$\rho$ the rest-mass density, $K=30000$ and $\Gamma=2.75$, in which
case the maximum gravitational mass is $M\simeq 2.43\,M_{\odot}$ for a
non-rotating NS, and $M\simeq 2.95\,M_{\odot}$ for a uniformly
maximally-rotating star, in agreement with recent observations of NS
masses~\citep{Demorest2010,Antoniadis2013}. An ideal-fluid EOS with
$\Gamma=2.75$ is used during the evolution in order to allow for shock
heating during merger.\footnote{We note that, if no shocks are
  present, the ideal-fluid and the polytropic EOSs are identical (see
  also~\citealt{Baiotti08}).} This EOS has been chosen since it fits
very well the Shen nuclear EOS~\citep{Shen1998a,Shen1998b} at high
densities~\citep{Oechslin2007}, and hence it provides a more accurate
description of the evolution of the plasma in the high-density regions
than the simpler $\Gamma=2$ polytrope used in our previous
simulations~\citep{Giacomazzo:2009mp, Giacomazzo2011,
  2011ApJ...732L...6R}. In this work we consider an equal-mass system
both with and without a magnetic field, and with a total gravitational
mass $M_{\rm ADM}=2.42M_{\odot}$. When a magnetic field is present,
its initial configuration is purely poloidal and aligned with the
angular momentum of the binary as in~\citet{Giacomazzo2011}. Details
about the initial configurations are provided in Table~\ref{table1}.

\section{Dynamics}
\label{dynamics}
\begin{figure*}[t]
  \centering
  \begin{tabular}{cc}
    \includegraphics[width=.4\textwidth]{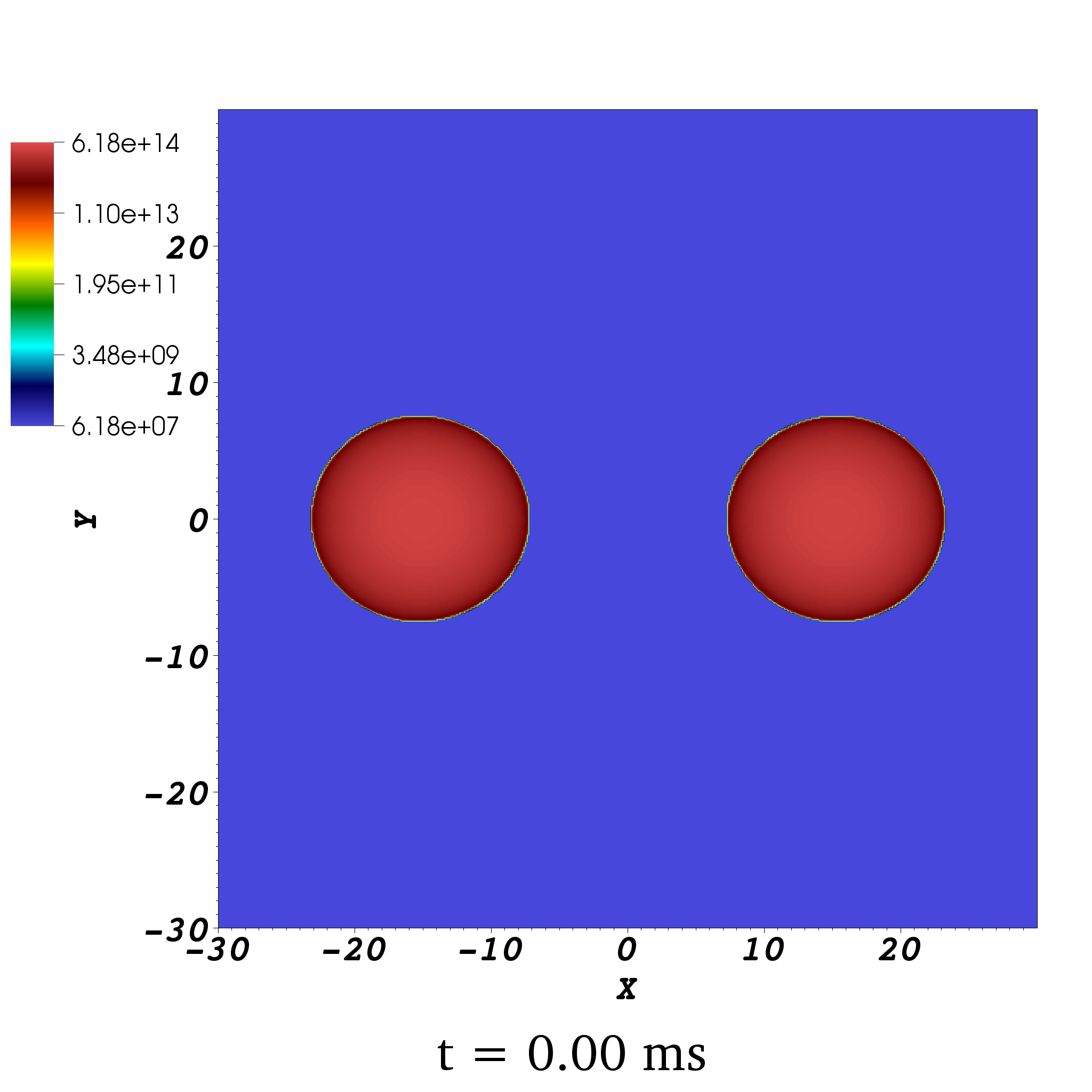}
    \includegraphics[width=.4\textwidth]{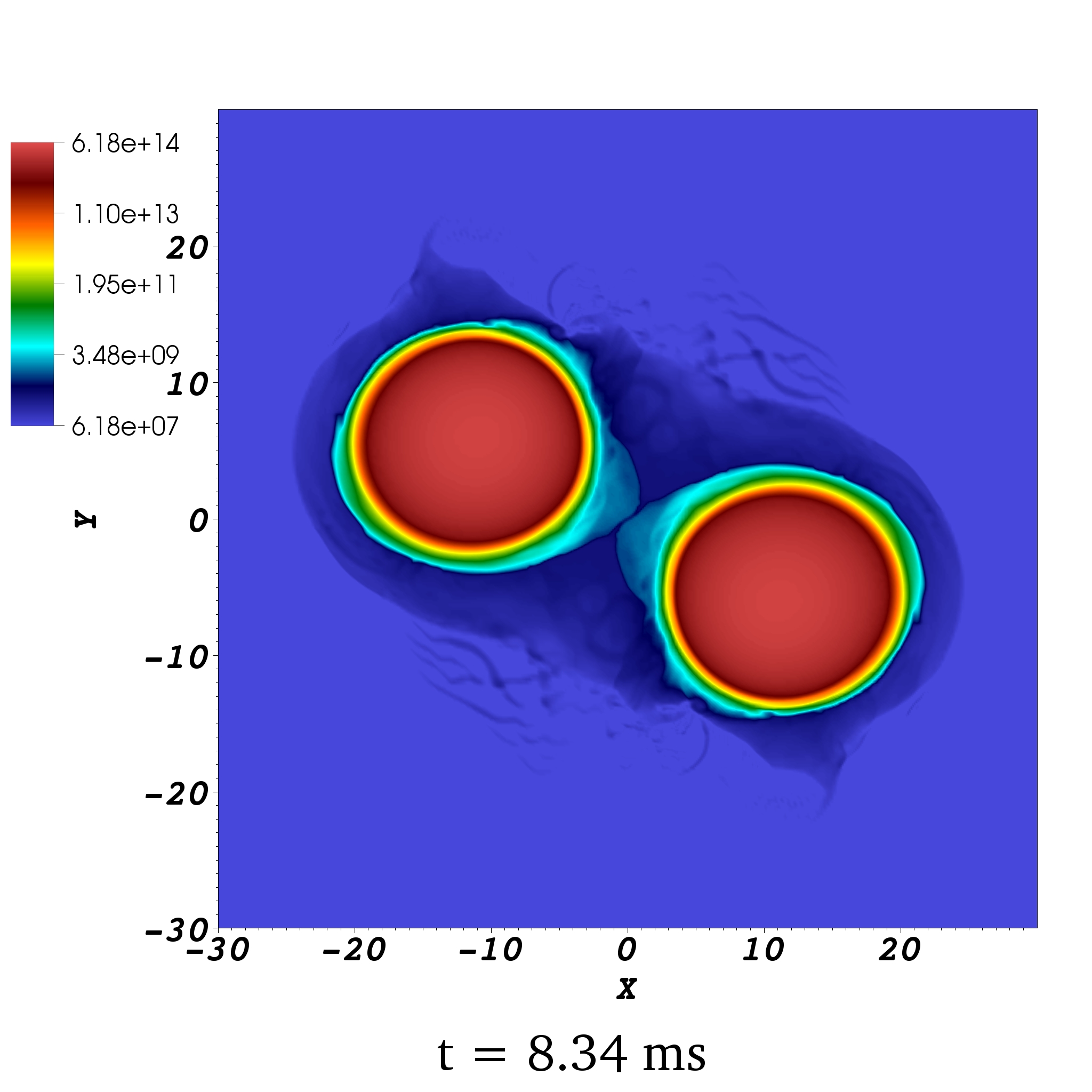}\\
    \includegraphics[width=.4\textwidth]{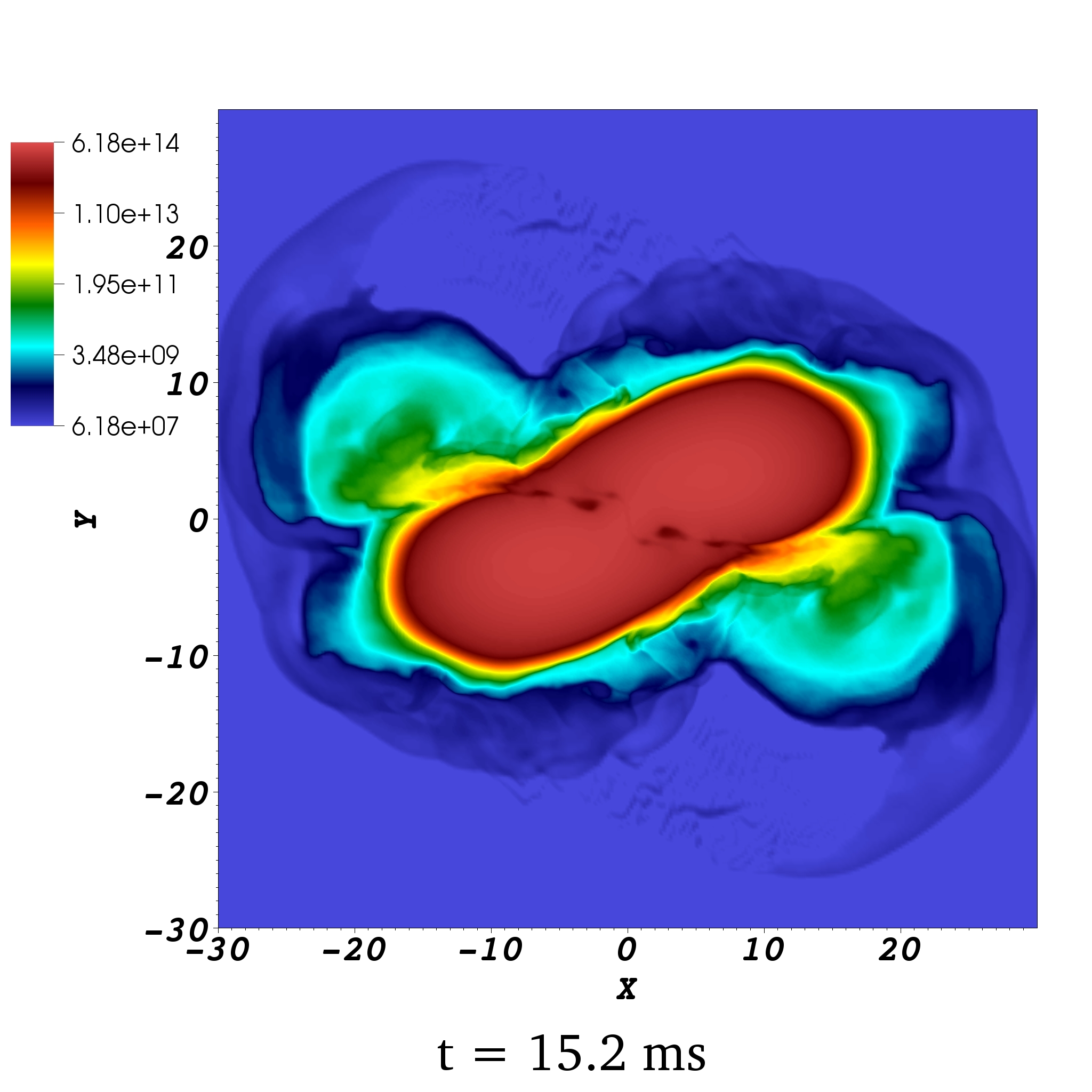}
    \includegraphics[width=.4\textwidth]{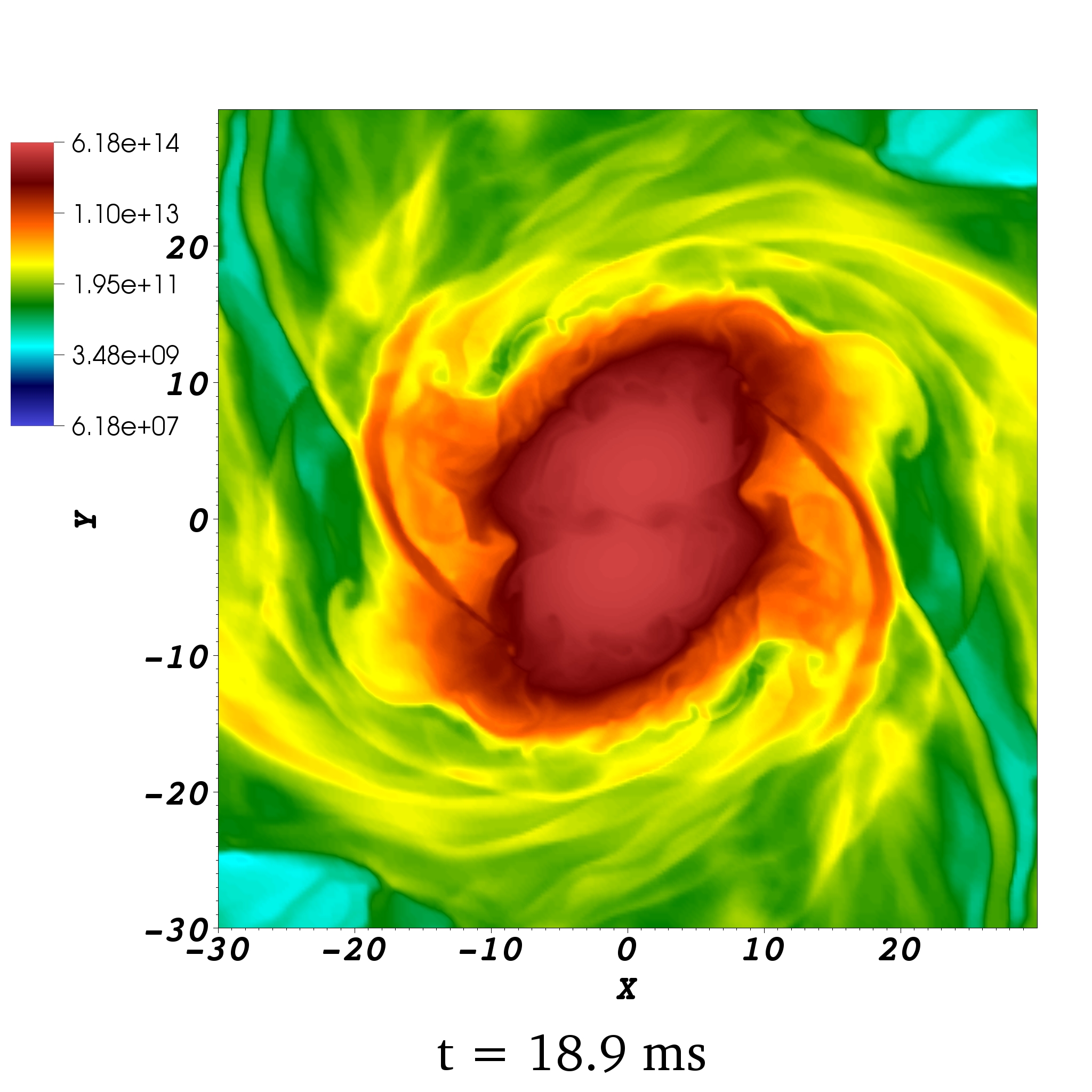}\\
    \includegraphics[width=.4\textwidth]{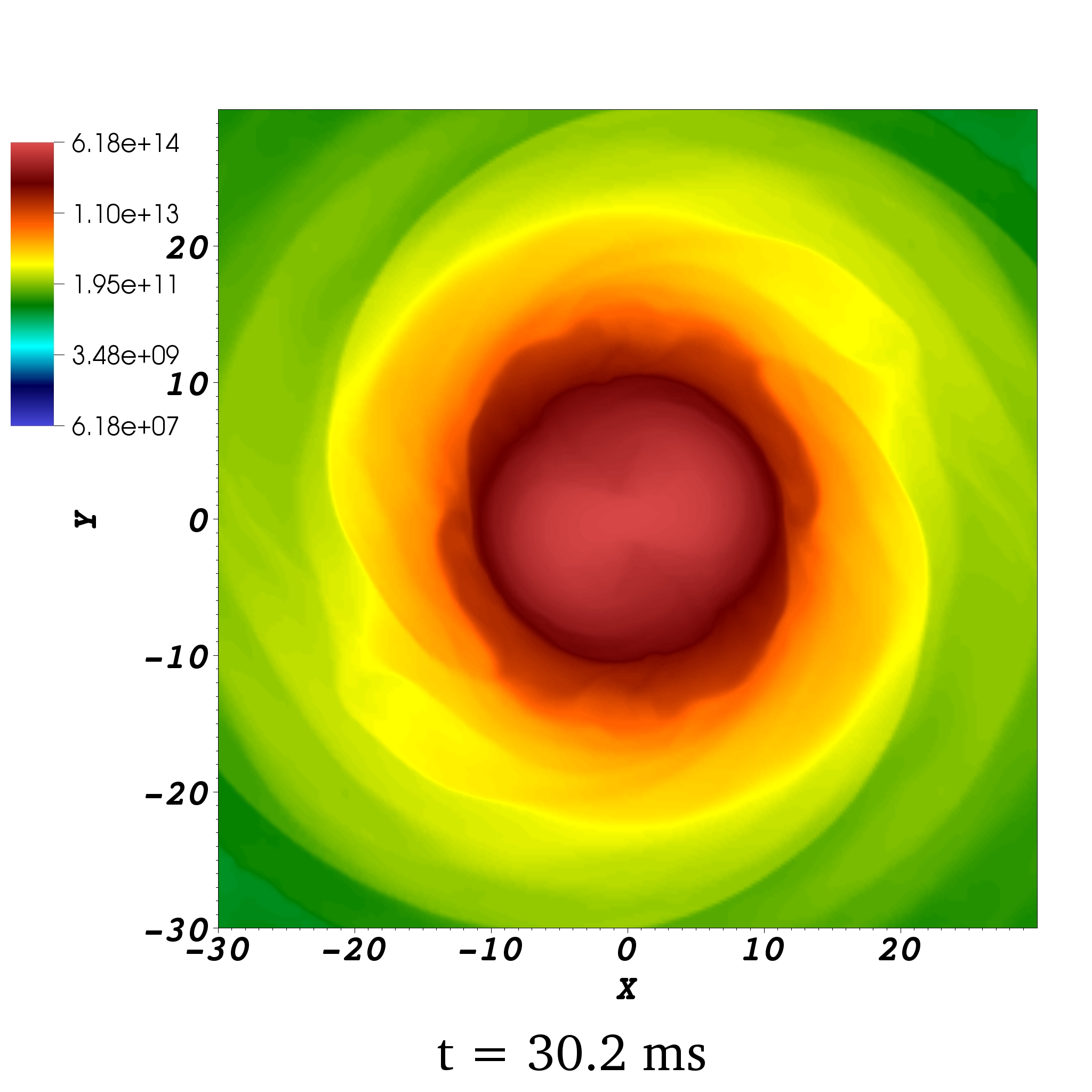}
    \includegraphics[width=.4\textwidth]{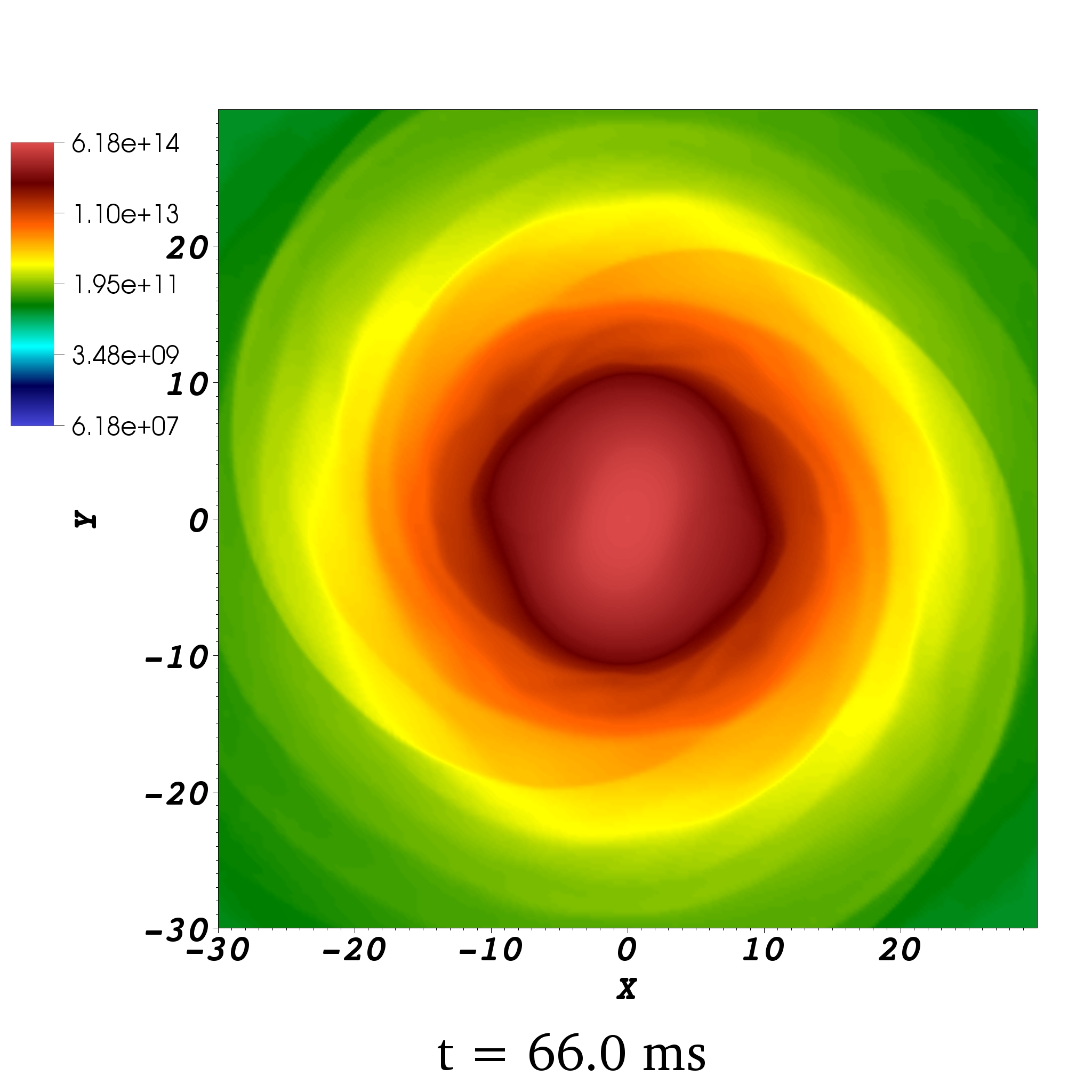}\\
  \end{tabular}
  \caption{Evolution of the rest-mass density in ${\rm g\, cm^{-3}}$ on
      the equatorial plane for model B0 evolved with the highest
      resolution ($h\sim 180 {\rm m}$). The different panels show
      respectively the initial conditions ($t=0$), the inspiral, the
      time of the merger ($t\sim 15 \,{\rm ms}$), the post-merger
      phase, and the formation of the ``ultraspinning'' NS (last two
      panels). The units of distance are $M_\odot \sim 1.5\, {\rm km}$
      and the time in ${\rm ms}$ is reported at the bottom of each
      panel. \label{figure1}}
\end{figure*}

We first evolved the unmagnetized case (model B0) with three different
resolutions: $h=0.24 M_\odot\sim 360 \,{\rm m}$ (low), $h=0.15
M_\odot\sim 225 \,{\rm m}$ (medium), and $h=0.12 M_\odot\sim 180
\,{\rm m}$ (high). In all cases, the binary inspirals for five orbits
before merging. The main aspects of the dynamics are illustrated in
Figure~\ref{figure1}, where we show the rest-mass density on the
equatorial plane for the high resolution case.
One important point to note is that, as observed in previous
simulations of BNS mergers, the compact object that is formed after
the inspiral is differentially rotating. However, while in previous
simulations it was an HMNS, in the current simulations the NS is well
below that limit ($M\simeq 2.95\,M_{\odot}$ for our EOS), and hence it
does not collapse to BH. At the end of our simulation, the
differentially-rotating NS has a mass $M \sim 2.36 M_\odot$ and is
surrounded by a disk of $\approx 0.1 M_\odot$.

\begin{figure*}[t!]
  \centering
  \begin{tabular}{cc}
    \includegraphics[width=.3\textwidth]{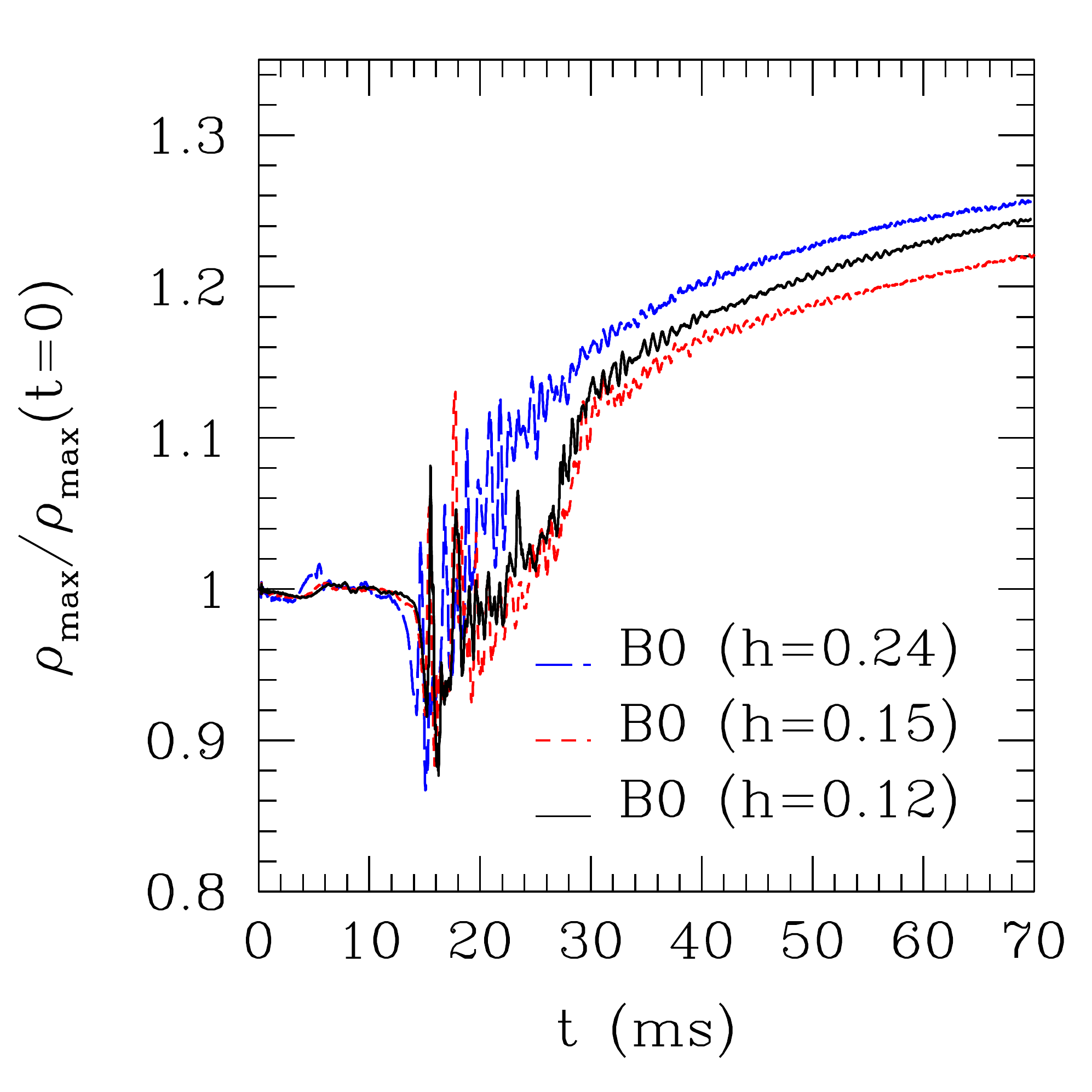}
    \includegraphics[width=.3\textwidth]{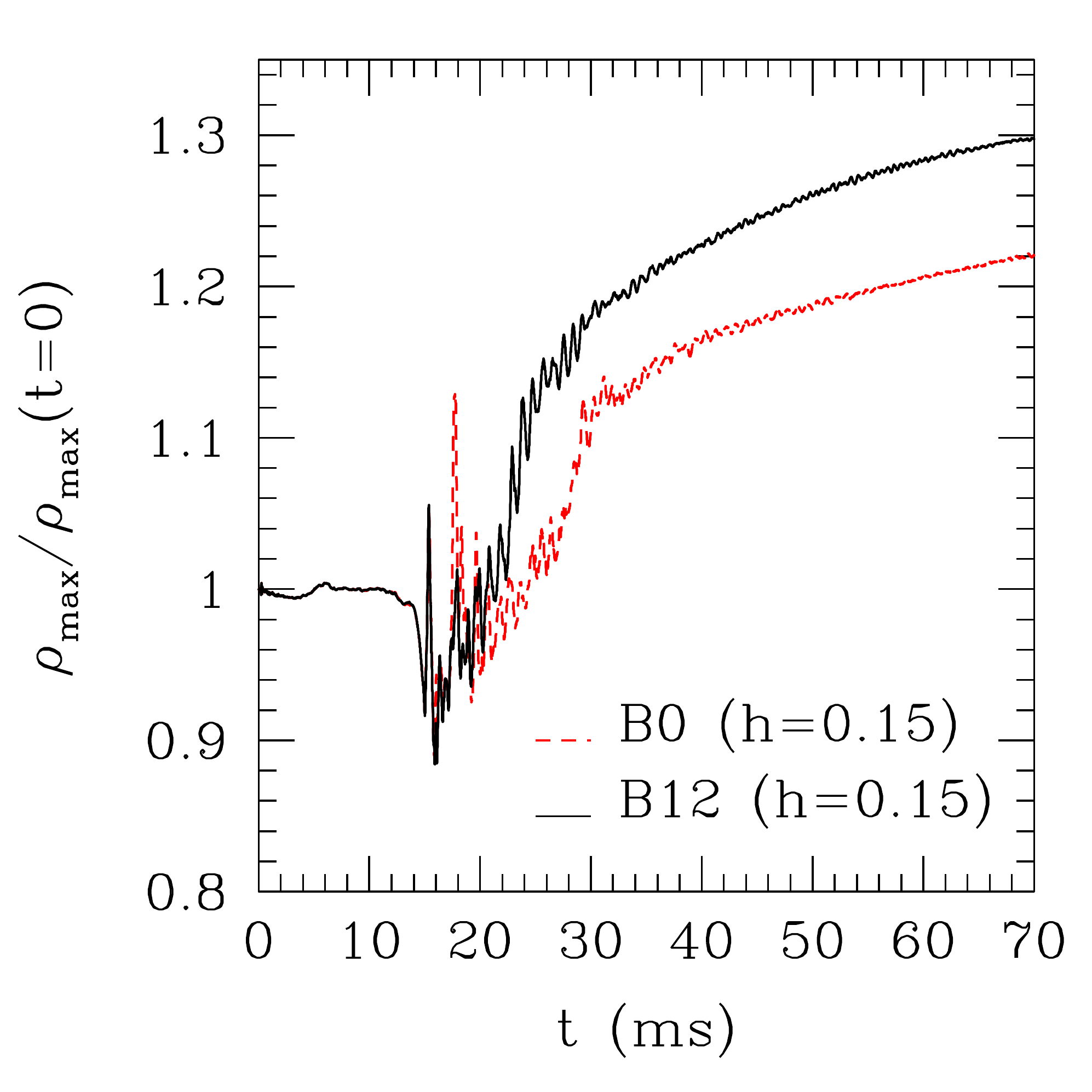}\\
    \includegraphics[width=.3\textwidth]{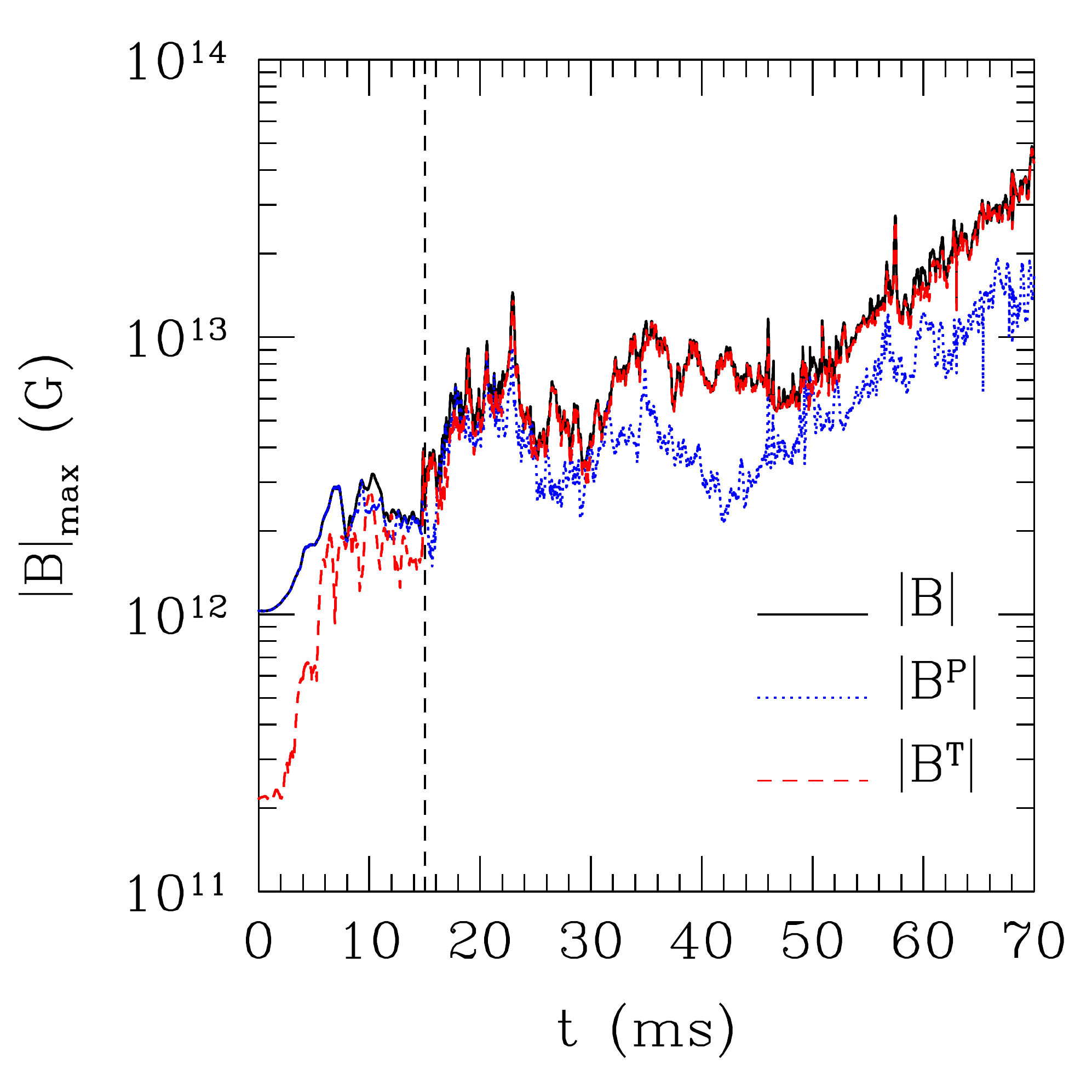}
    \includegraphics[width=.3\textwidth]{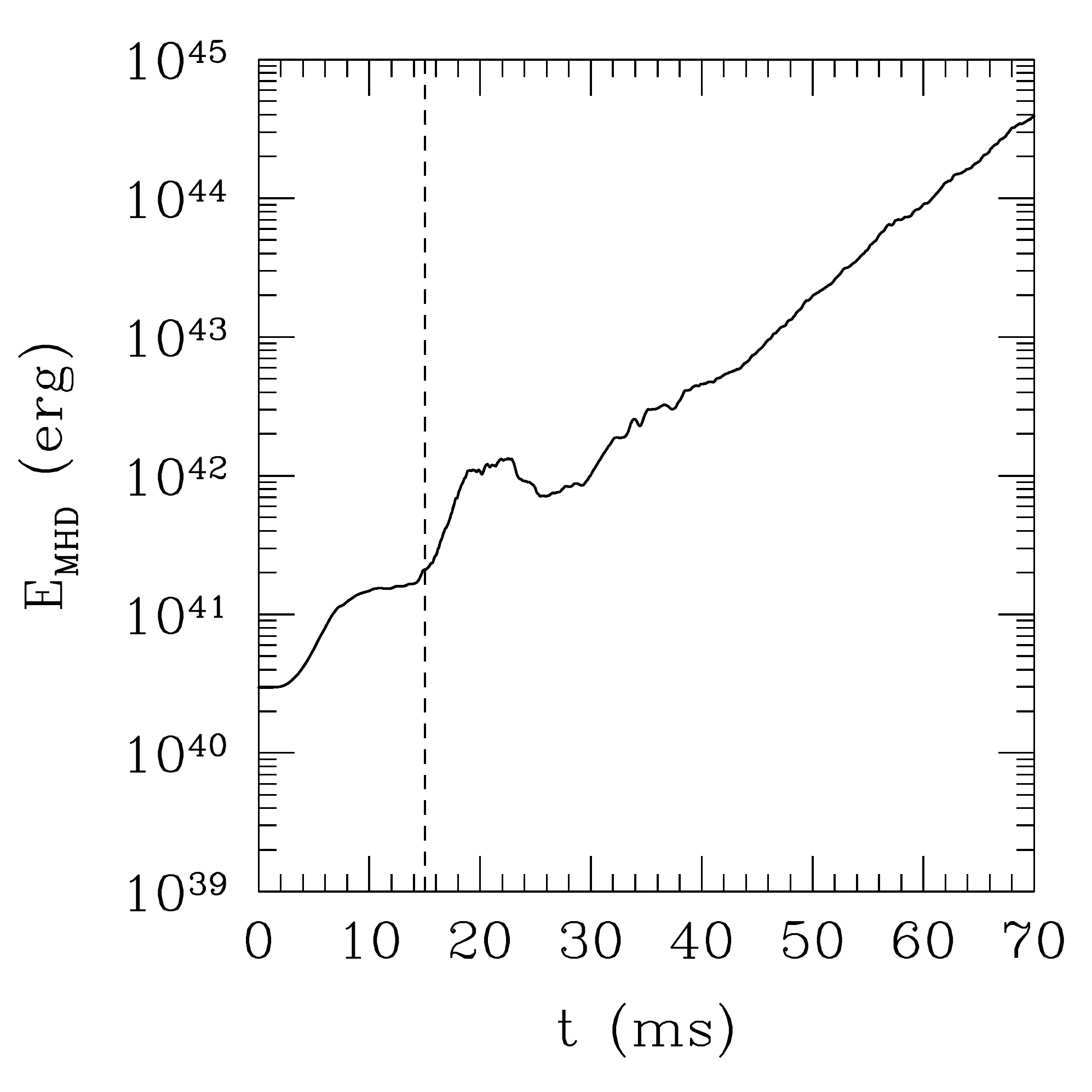}\\
  \end{tabular}
  \caption{Top left panel: evolution of the maximum of the rest-mass
    density $\rho$ normalized to its initial value for model B0
    evolved with three different resolutions. The blue long-dashed
    line refer to $h=0.24 M_\odot\sim 360 \,{\rm m}$, the red
    short-dashed line to $h=0.15 M_\odot\sim 225 \,{\rm m}$, and the
    black solid line to $h=0.12 M_\odot\sim 180\, {\rm m}$. Top right
    panel: same quantity as in the left panel, but for models B0 (red
    short-dashed line) and B12 (black solid line) evolved with our
    fiducial resolution ($h=0.15 M_\odot\sim 225 \,{\rm m}$). Bottom
    left panel: evolution of the maximum of the magnetic field for
    model B12. The black solid line shows the maximum of the total
    magnetic field, the blue dotted line the maximum of its poloidal
    component, and the red dashed line the maximum of its toroidal
    component. Bottom right panel: evolution of the total magnetic
    energy as a function of time. In the last two panels the vertical
    dashed line shows the time of the merger of the two NS cores.
    \label{figure2}}
\end{figure*}

This can also be seen in the top left panel of Figure~\ref{figure2}, where
we show the evolution of the maximum of the rest-mass density
normalized to its initial value for the three resolutions. Following
an initial transient after the merger at $t\sim 15 \,{\rm ms}$, when
the rest mass density increases by $\sim 15 \%$ due to the compression
of the NS cores, the rate of the rest-mass increase diminishes, and
the central density approaches a finite value. Simulations of unstable
HMNSs, on the other hand, show a clear increase in the rest-mass
density (see, e.g., the left panel of Figure A1
in~\citealt{2010CQGra..27k4105R}). 

Since these objects are differentially rotating NSs, it is interesting
to further explore what happens when magnetic fields are added to the
simulation, as they may redistribute angular
momentum~\citep{Giacomazzo2011}. The top right panel of
Figure~\ref{figure2} shows a comparison between the evolution of the
maximum of the rest-mass density for models B0 and B12. While model
B12 shows a larger increase after the merger, due to redistribution of
angular momentum by the magnetic field, its maximum density is also
converging toward a constant value. In the bottom left panel of
Figure~\ref{figure2}, we show the evolution of the maximum of the
magnetic field for model B12. As already observed in our previous
simulations~\citep{Giacomazzo2011}, the magnetic field grows by one
order of magnitude during the merger because of hydrodynamic
instabilities such as the Kelvin-Helmholtz (KH; see
also~\citealt{Price2006,Baiotti08}), and because of compression of the
cores. The KH also causes an increase of the toroidal component of the
magnetic field. This component is further amplified by magnetic
winding due to differential rotation. The amplification of the
magnetic field can also be observed in the bottom right panel of
Figure~\ref{figure2}, where we show the total magnetic energy $E_{\rm
  MHD}$ as a function of time. At the end of the simulation, $E_{\rm
  MHD}\sim 10^{44}\, {\rm erg}$, but the $E_{\rm MHD}(t)$ slope
suggests that further growth is expected if the simulation were
continued for a longer time. We also note that local simulations show
that the KH instability can further amplify the magnetic field up to
$\sim 10^{16} {\rm G}$, i.e, well into magnetar
levels~\citep{Zrake2013}. Unfortunately, the resolutions required to
correctly resolve such small scale dynamics and other instabilities,
such as the magnetorotational instability (MRI), are far above what
can be obtained in these global simulations.
We also note that differential rotation can cause a further increase
of the magnetic field at later times~\citep{Thompson1993}
and hence magnetar field levels could be reached via several
mechanisms.

\begin{figure}[t!]
  \centering
  \begin{tabular}{c}
    \includegraphics[width=.4\textwidth]{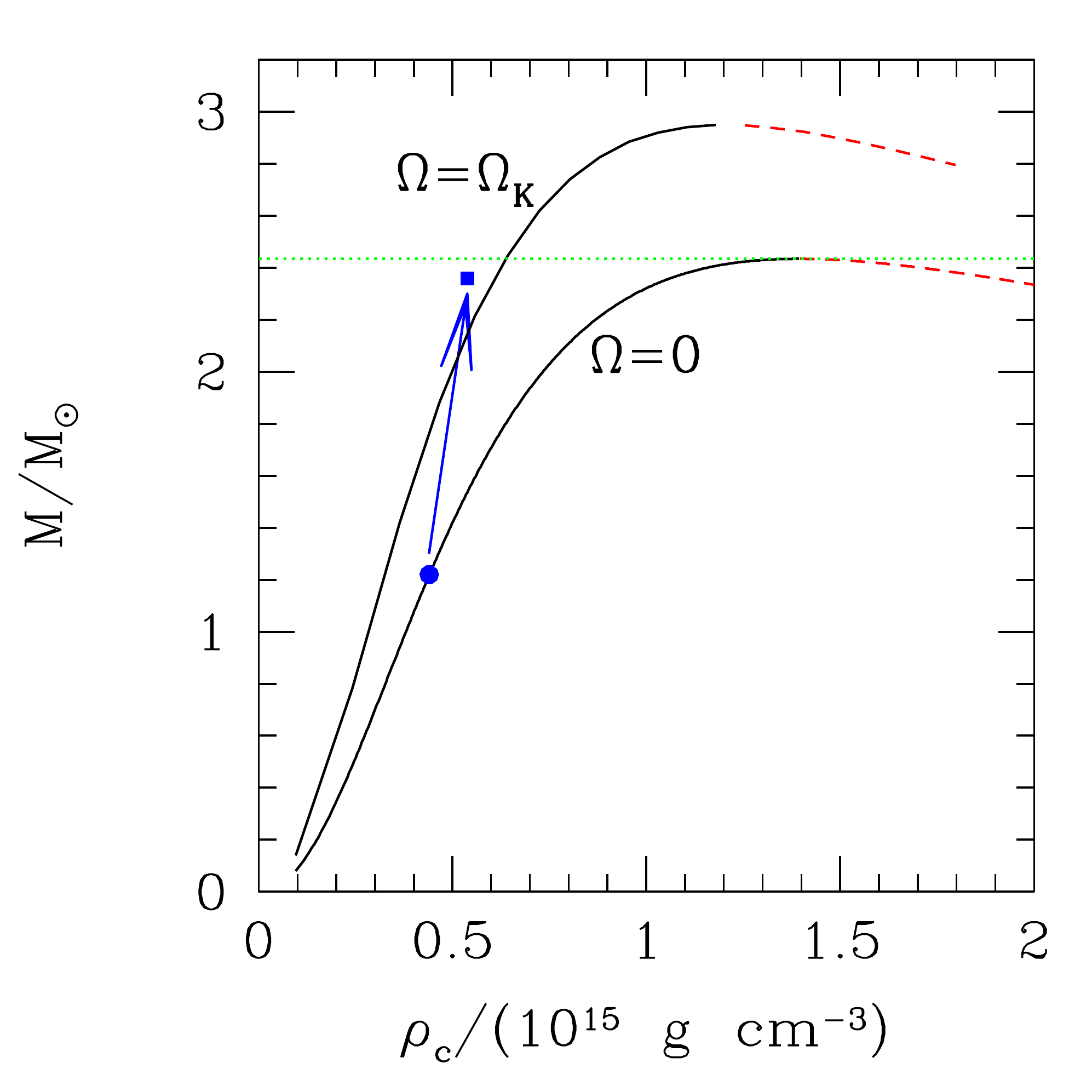}
  \end{tabular}
  \caption{The gravitational mass of an NS as a function of the
    central value of its rest-mass density $\rho_c$. The solid
      and dashed lines represent equilibrium solutions for uniformly
      rotating NSs. The bottom black solid line refers to stable
      non-rotating NSs (i.e, TOVs), while the bottom red dashed line
      to gravitationally unstable non-rotating NSs (note that they have
      masses below the maximum mass). The top black-solid and
      red-dashed lines refer respectively to stable and unstable NSs
      rotating at the mass shedding limit. Uniformly rotating NSs
      located in the region between the two red dashed lines are
      unstable and will collapse to BH. The filled blue circle shows
    the position of the NSs composing our binary, while the filled
    blue square indicates the NS formed at the end of the simulation
    of model B0. The horizontal green dotted line shows the maximum
    mass for a non-rotating NS. As one can easily see, the NS formed
    after the merger has a mass lower than the maximum mass for a
    non-rotating NS and it is located in the stable region.
    \label{figure3}}
\end{figure}

For a clearer interpretation of our results, in Figure~\ref{figure3}
we show equilibrium curves for uniformly rotating NSs. The two
  black solid lines show non-rotating and maximally-rotating NSs that
  are stable against gravitational collapse, while 
  the two red-dashed lines straddle the region of uniformly rotating
  models which are unstable. NSs with densities larger than $\sim
  1.2 \times 10^{15}\,{\rm g\, cm^{-3}}$ are unstable and
  collapse to BHs (this is true also for differentially rotating NSs,
  see~\citealt{diffrot}). The filled blue circle shows the position
of the two NSs composing our binary, while the blue square indicates
the position of the NS produced by the merger at the end of the
simulation, for model B0 at medium resolution. First, as
  mentioned before, its mass and central density are lower than the
  maximum value for a stable uniformly-rotating NS, hence it does not form
  an HMNS. This object is differentially rotating with spin parameter
$J/M^2\sim0.86$; its total angular momentum $J$ is higher, by a factor
$J/J_{\rm max}\sim 1.11$, than the maximum angular momentum $J_{\rm
  max}$ which can be obtained for a rigidly rotating NS with the same
rest mass. This is why it is located above the top black
curve. 

\citet{Duez2006} studied the evolution of several stable
differentially rotating and magnetized NS models in two dimensions and
at much higher resolutions than what can be afforded in
three-dimensional simulations of BNS mergers. One of their models,
which they call ``ultraspinning'', is very similar to the end product
of our BNS merger simulations. In particular, \citet{Duez2006} studied
the impact of the magnetic field on the long-term evolution of this
model. Their simulations showed an amplification of the magnetic field
due to the onset of the MRI instability, and found that the final
configuration was an uniformly rotating NS surrounded by a
differentially rotating and magnetized disk.

\section{Gravitational Waves and Short Gamma-Ray Bursts}
\label{GWs}
Here we discuss the impact of our simulations for the detection of GW
signals, as well as for the possible connection with current
observations of SGRBs.

\begin{figure}[t!]
  \centering
  \begin{tabular}{cc}
    \includegraphics[width=.4\textwidth]{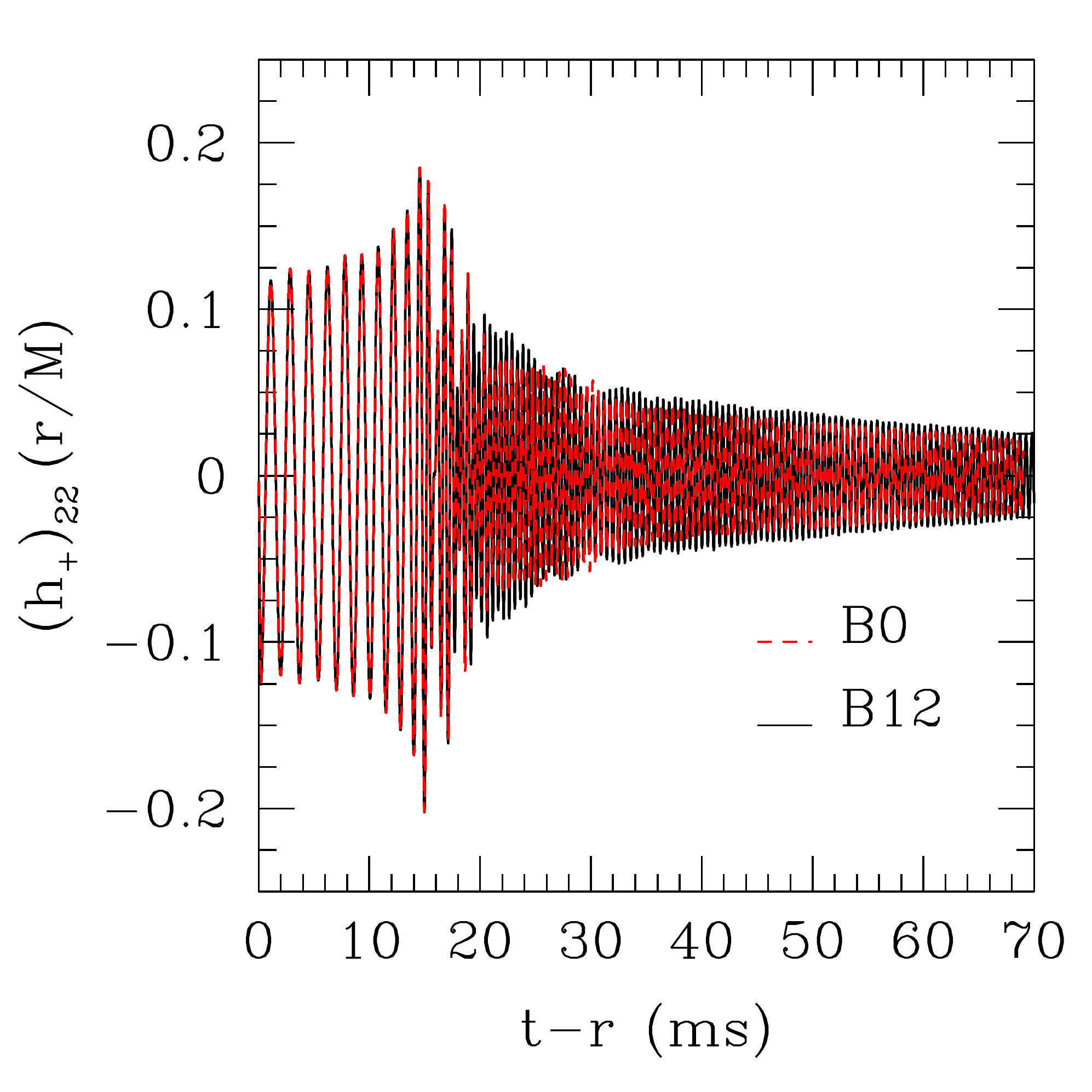}
  \end{tabular}
  \caption{The $l=2,m=2$ mode of the GW signal for model B0 (red
    dashed line) and B12 (black solid line).
    \label{figure4}}
\end{figure}

Figure~\ref{figure4} shows the amplitude of the $l=2,m=2$ mode of the
GW signal for our models B0 and B12, run at medium resolution. As
already observed in~\citet{Giacomazzo:2009mp} and
in~\citet{Giacomazzo2011}, the magnetic field does not have an impact
on the inspiral, but it affects the signal after the merger. In
particular, due to the slightly larger compactness of the magnetized
NS, the GW signal is slightly larger in amplitude.

As already mentioned in the previous section, long-term simulations of
``ultraspinning'', magnetized, differentially rotating NSs (similar to
the ones produced at the end of our BNS merger simulations) have shown
that the end product of the evolution is an uniformly-rotating NS
surrounded by a differentially rotating magnetized
disk~\citep{Duez2006}. In addition, those simulations were able to
resolve the MRI and showed the formation of a mainly poloidal and
collimated magnetic field aligned with the spin axis of the NS. Such a
configuration could emit relativistic jets and power
SGRBs~\citep{Meier2001}. This possibility is especially interesting in
light of the recent observations of extended emission following
SGRBs~\citep{Metzger2008}. An analysis of {\em Swift}-detected SGRBs
by \citet{Rowlinson2013} has showed that all SGRBs with one or more
breaks in their X-ray light curves display a plateau phase, which can
be interpreted as the luminosity of a relativistic magnetar
wind~\citep{zhang2001,fan06,Metzger2011}. Under the assumption of
energy loss by pure dipole radiation, and neglecting, to first
approximation, the enhanced angular momentum losses due to
neutrino-driven mass loss, the duration of the plateau and its
luminosity can be used to infer the magnetic field of the magnetar and
its birth period. The observed range of values (plateau
  durations $\sim 10^2-10^4$~s, and [$1-10^4$~keV] luminosities $\sim
  10^{46}-10^{49}$~erg~s$^{-1}$) yielded typical periods on the order
of a few milliseconds, and magnetic field strengths in the range
$B\sim 10^{15}-10^{16}$~G. Following the initial rapidly spinning
magnetar phase, two outcomes are possible, depending on how steep the
post-plateau decay phase is. If the magnetar is unstable and decays to
a BH, then the post plateau emission, only due to curvature radiation,
fades away very quickly. On the other hand, the $\sim t^{-2}$ decay of
the stable magnetar emission gives a more prolonged energy injection,
and hence brighter fluxes at later times. The detailed analysis by
\citet{Rowlinson2013} identified a handful of SGRBs whose late X-ray
emission is consistent with that of a stable magnetar. Moreover,
  X-ray and optical afterglow emitted by a
  magnetar~\citep{Dallosso2011,Zhang2013} may not be 
  collimated, and hence they may be observed even without a SGRB
  detection~\citep{Gao2013}.

Other numerical simulations of magnetized HMNSs have further demonstrated 
the possibility of producing outflows with energy of $\sim 10^{51}
{\rm erg}$ for magnetic fields of $\sim 10^{15} {\rm
  G}$~\citep{Kiuchi2012}. As already discussed before, such magnetic
fields can be naturally formed in our scenario via KH and MRI
instabilities. According to~\citet{Kiuchi2012}, a magnetic field of
$\sim 10^{15}{\rm G}$ could give rise to an electromagnetic emission
observable in the radio band and hence provide an interesting
electromagnetic counterpart to the GW signal even if a SGRB is not
observed.

\section{Summary}
\label{conclusions}
We have presented the first general relativistic magnetohydrodynamic
simulations that show the possible formation of a stable magnetar. The
NS formed after the merger is found to be differentially rotating and
ultraspinning. Since our computational resources are not enough to
fully resolve the MRI, the magnetic field is amplified by about two
orders of magnitude, but further amplification is possible and
indeed observed in two and three-dimensional simulations of
differentially rotating NSs~\citep{Duez2006,Siegel2013}. Moreover,
long term evolution of such models has shown that the magnetic field
can impact the angular velocity profile of the NS leading to the
formation of an uniformly rotating NS surrounded by an accretion disk
and with a collimated magnetic field~\citep{Duez2006}. While it will
be difficult to differentiate the GW signal between the magnetized and
the unmagnetized scenarios, strong electromagnetic counterparts that
would be suppressed in collapsing NSs could be easily produced and
observed in radio~\citep{Kiuchi2012},
optical~\citep{Dallosso2011,Zhang2013,Gao2013},
X-rays~\citep{Rowlinson2013}, and gamma-rays~\citep{Gompertz2013}.

While our simulations focused on equal-mass systems, the same scenario
may be produced after the merger of unequal-mass BNSs. In this case,
matter ejected during the inspiral due to the tidal disruption of the
less massive components, may later fall back on the magnetar and
trigger its collapse to BH~\citep{Giacomazzo2012b}.  More detailed
observations of the early afterglow phase, as expected with the
planned future mission {\em LOFT}~\citep{Amati2013}, will be
especially useful in discriminating among various formation
scenarios. Last, simultaneous detections of GWs and SGRBs will fully
unveil the mechanism behind the central engine and help constrain its
properties~\citep{Giacomazzo2013}.

\acknowledgments We thank Brian Metzger, Christian Ott, Antonia
Rowlinson, Luigi Stella, Eleonora Troja, Bing Zhang and an anonymous
referee for useful comments.  B.G. and R.P. acknowledge support from
NSF grant No. AST 1009396 and NASA grant No. NNX12AO67G. This work
used XSEDE (allocation TG-PHY110027) which is supported by NSF grant
No. OCI-1053575.


\bibliographystyle{apj}


\end{document}